# Medical Image Quality Metrics for Foveated Model Observers

Miguel A. Lago,[a] Craig K. Abbey,[a] Miguel P. Eckstein[a,b*]
[a] Department of Psychological and Brain Sciences, University of California at Santa Barbara, Santa Barbara, CA 93106 USA
[b] Department of Electrical and Computer Engineering, University of California at Santa Barbara, Santa Barbara, CA 93106 USA

**Abstract**

**Purpose**: A recently proposed model observer mimics the foveated nature of the human visual system by processing the entire image with varying spatial detail, executing eye movements and scrolling through slices. The model can predict how human search performance changes with signal type and modality (2D vs. 3D), yet its implementation is computationally expensive and time-consuming. Here, we evaluate various image quality metrics using extensions of the classic index of detectability expressions and assess foveated model observers for location-known exactly tasks.

**Approach**: We evaluated foveated extensions of a Channelized Hotelling and Non-prewhitening model with an eye filter. The proposed methods involve calculating a model index of detectability (d') for each retinal eccentricity and combining these with a weighting function into a single detectability metric. We assessed different versions of the weighting function that varied in the required measurements of the human observers' search (no measurements, eye movement patterns, and size of the image and median search times).

**Results**: We show that the index of detectability across eccentricities weighted using the eye movement patterns of observers best predicted human performance in 2D vs. 3D search performance for a small microcalcification-like signal and a larger mass-like. The metric with weighting function based on median search times was the second best at predicting human results.

**Conclusions**: The findings provide a set of model observer tools to evaluate image quality in the early stages of imaging system evaluation or design without implementing the more computationally complex foveated search model.

*Third Author, E-mail: eckstein@psych.ucsb.edu

1. Introduction

For over four decades, researchers have worked on metrics of medical image quality. Some efforts have concentrated on physical measures of image quality that consider the relationship between the signal, noise, and the imaging system's modulation transfer function [1]. Researchers increasingly sought to incorporate the human visual system's properties to capture a radiologist's performance, visually detecting and classifying lesions. These mathematical formulations are known as model observers, which can be applied to statistical properties of the signals and backgrounds or the actual images and result in a performance measure detecting the signal [1]–[11]. Through four decades, the field of medical imaging has made progress in the development and evaluation of model observers. Early work concentrated on computer-generated backgrounds and simple detection tasks with a single signal appearing at one or a few locations (location-known exactly, LKE) [4], [6], [10]–[14]. With the advent of large scale digitalization of medical images in the mid-1990's model observers were first applied to large samples of anatomical backgrounds extracted from x-ray coronary angiograms and mammograms [15]–[22]. Further developments included signals that could vary in size and shape [10], [20], [22]–[25], dynamic and/or 3D components [3], [26]–[33], and incorporation of properties of the human visual system including spatial frequency channels, internal noise, non-linear transducers and divisive normalization [11], [13], [34].

Still, a majority of studies have evaluated the validity of model observers with simple detection tasks for which the target might appear in a single or few known locations within the image [10], [20], [23], [30], [35]–[47]. The underlying assumption is that model observer performance for simple visual detection tasks with a few known locations reliably predicts performance with more clinically realistic tasks, which involve searching for the lesion across a larger region in the image. If so, conclusions from evaluations and optimizations of imaging systems with simple LKE tasks will be applicable and valid in more clinically realistic search tasks.

For search in 2D images, researchers have used signal detection theory to quantitatively map performance across tasks with varying numbers of known possible signal locations in 2D images [12], [15], [48]–[50]. Investigators have also demonstrated shortcomings of the simple location-known exactly detection tasks and motivated the need to develop model observers that account for the visual search process [51]–[55]. Studies have also found discrepancies between the rank ordering of imaging systems based on LKE and search tasks [56], [57].

The dissociation between LKE tasks and search tasks becomes even more pronounced with 3D images. When searching with 3D images, observers typically do not exhaustively scrutinize through eye movement every region of each slice with the central region of the eye, the fovea. Such foveal region can process visual information with high spatial detail [58]–[60]. Instead, observers rely on processing much of the 3D image data with the visual periphery with lower spatial acuity. In contrast, for LKE tasks or even 2D search, observers scrutinize all possible target locations with the fovea. Studies have shown how the difference in how observers visually process information in LKE, 2D, and 3D images leads to dissociation in human performance [61].

Lago et al. [62], [63], have shown that, for 3D images, conventional model observers [3], [5], [8], [10], [11], [13], [16], [27], [30], [35], [38], [42], [45], [64] that do not take into account the degradation in processing in the human visual periphery are unable to predict detriments in performance for small signals in 3D search. They proposed a model (foveated search model) that combines elements of model observers for medical images and those of search from vision science [65]–[70]. The model processes the visual field with varying spatial detail based on the current fixation (point of gaze of the model), makes eye movements, scrolls through the slices, and integrates information across the search process to reach target present/absent decisions. The advantages of the model are that, unlike traditional model observers, it mimics humans in how they scrutinize images capturing aspects of human behavior such as explicit saccades and scrolls. This model is also able to predict the source of errors (search vs. recognition errors).

The main disadvantage is that the model is computationally intensive and difficult to compute. It requires N times more dot products than a scanning observer [51], [52], [71], depending on the number of templates for the foveated model (N) and also some complementary information about the extent of the exploratory eye movement behavior of humans (e.g., percentage of the volume explored).

Thus, our current goal is to investigate several possible simplifications using traditional calculations for signal detectability for signal known exactly/location known exactly [1], [2], [8], [42], [64] and relate them to the foveated search model's predictions for 2D and 3D search. The approach requires a model that can predict humans' ability to detect the signal as a function of retinal eccentricity. If successful, these image quality metrics based on detectability calculations should provide a first pass approximation to the previously proposed full foveated search model (FSM). The metric s could potentially be used as a proxy to evaluate and optimize image quality without implementing the cumbersome FSM.

We first present the theory and figures of merit based on the index of detectability for a SKE/LKE and their extensions for a foveated model. We then assess whether the figures of merit for the foveated model can predict human performance during search in with 2D and 3D search in $1/f^{2.8}$ filtered noise. In particular, we evaluate whether the index of detectability metrics for the foveated model (based on an extension of the Channelized Hotelling Observer [14], [26], [34], [35], [39], [40] and the Non-Prewhitening Model with Eye Filter [5], [10], [13], [45]) can correctly predict human: 1. the detriment of 3D search for a small sharp-edged signal that mimics a microcalcification (when compared to 2D search); 2. No detrimental influence of 3D images on performance for a larger mass-like signal.

## 2. Theory

### 2.1 Model Observer Basics

A majority of model observers are evaluated by taking the dot product between a pre-defined model template **w** and data $\mathbf{g}_i$ at the possible target locations $i$ to result in a scalar response $\lambda_i$ used to make decisions [1], [5], [8], [39], [72]:

$$\lambda_i = \mathbf{w}^t \mathbf{g}_i \tag{1}$$

Models vary in the assumed templates which can take into account signal luminance properties [2], [10], [13], [41], [44], [45], noise statistical properties [4], [5], [73], [74], and constraints of the human visual system [14], [34], [72], [75].

To quantify search performance in a detection task, the decision variable ($\lambda$) of the model can be evaluated for target-present and target-absent trials to compute the hit rate, false-positive rate, and an area under the ROC curve by varying the decision threshold.

Often times, when the decision variable is Gaussian distributed, performance is also quantified using a $d'$ index of detectability, which quantifies the difference in response to the signal vs. noise in standard deviation units [4], [64], [76]:

$$d' = \frac{\lambda_s - \lambda_n}{\sigma_\lambda} \quad (2)$$

When the model observer is linear, then, the index of detectability can be directly calculated from the template $\mathbf{w}$, the signal luminance profiles, and the covariance matrix $\mathbf{K}$.

$$d' = \frac{\mathbf{w}^ت \mathbf{s}}{\sqrt{\mathbf{w}^ت \mathbf{K} \mathbf{w}}} \quad (3)$$

where the superscript t denotes transpose.

Furthermore, when the noise is statistically stationary (the variance and covariance does not depend on location), $d'$ can be estimated from the Fourier domain:

$$d' = \frac{\int_{-\infty}^{+\infty} \int_{-\infty}^{+\infty} \overline{w(u,v)} s(u,v) \, du \, dv}{\int_{-\infty}^{+\infty} \int_{-\infty}^{+\infty} \sqrt{|w(u,v)|^2 N(u,v)} \, du \, dv} \quad (4)$$

where $w(u,v)$ is the template, $s(u,v)$ is the signal luminance, and $N(u,v)$ is the noise power spectrum. $|w(u,v)|$ stands for the absolute value of the template and $\overline{w(u,v)}$ stands for the complex conjugate of the template.

## 2.2 Figures of Merit for Foveated Model Observers

### 2.2.1 Index of detectability for each retinal eccentricity

The classic theoretical treatment of model observers, described in the previous section, considers a single template representing the central vision of the human observer. However, for many new applications, such as 3D images, the observer typically utilizes other parts of their visual field to scrutinize image data. A first step is to characterize a performance metric for different parts of the visual field. Instead of defining a single $d'$, index of detectability, image quality can be characterized by a family of detectabilities spanning the visual field. As a first approximation, we assume that the deterioration in visual quality is anisotropic (rotationally invariant) as a function of distance from the fovea and represented by a distance of the signal location in visual degrees from central vision. The index of detectability can then be extended to $d'_E$:

$$d'_E = \frac{\lambda_{s,E} - \lambda_{n,E}}{\sigma_{\lambda_E}} \tag{5}$$

Each of the indices of detectability $d'_E$ is calculated from the decision variables ($\lambda$) with each corresponding perceptual template for a given eccentricity $\mathbf{w}_E$. The result is a collection of $d'$s spanning different retinal eccentricities. Thus, equation 3 can be extended to:

$$d'_E = \frac{\mathbf{w}_E^t \mathbf{s}}{\sqrt{\mathbf{w}_E^t \mathbf{K}_E \mathbf{w}_E}} \tag{6}$$

For statistically stationary noise, the Fourier expression of equation 4 can be generalized for different eccentricities and associated templates:

$$d'_E = \frac{\int_{-\infty}^{+\infty}\int_{-\infty}^{+\infty} \overline{w_E(u,v)} s(u,v)\, du\, dv}{\int_{-\infty}^{+\infty}\int_{-\infty}^{+\infty} \sqrt{|w_E(u,v)|^2 N(u,v)} du dv} \tag{7}$$

where the terms are defined below equation 4.

### 2.2.2 Integrating detectabilities across eccentricities

The previous section detailed how to calculate an index of detectability for each retinal eccentricity. The next step is to aggregate these indices into a single measure that can be used as a figure of merit to evaluate or optimize image quality for search. We can compute a weighted average across eccentricities:

$$\langle d' \rangle = \sum_{E=0}^{e} m_E d'_E \tag{8}$$

where $m_E$ is a weight assigned to the $d'_E$ at different eccentricities, and the summation extends from the fovea ($E = 0$) to the largest eccentricity in the image ($e$).

#### 2.2.2.1 Average d' across eccentricities

One possibility is trying to optimize a simple average ($m_E = 1$ for all eccentricities) of all the $d'$s across the entire visual field. This strategy would give equal importance to visual processing at all points in the visual field. Yet, it would not reflect the well-known fact that humans make three eye movements per second to fixate objects of interest [77], [78] and, even though they are able to process information in parallel [79], they predominantly use foveal information to make perceptual decisions.

#### 2.2.2.2 d' weighted average across eccentricities

A second possibility is to use a weighting function that gives more prevalence to the foveal region and proportional to the normalized detectability ($\sum_E \text{norm}(d'_E) = 1$) at a given eccentricity, $m_E = \text{norm}(d'_E)$. In this case, the optimization would be over:

$$\langle d' \rangle = \sum_{E=0}^{e} \text{norm}(d'_E) d'_E \tag{9}$$

#### 2.2.2.3 Frequency of minimum weighted $d'_E$ estimated from eye movements measurements (ET closest fix)

Arguably, the most accurate approach would be to have an estimate of the probability that the observer uses different parts of the visual field to process the signal during the process of visual search on each trial.

Thus, the $d'$ on each trial will vary depending on the specific eye movements and the various eccentricities of the signal across the multiple fixations.

For simplicity, we assume that the maximum attainable $d'$ for a trial is given by the minimum retinal eccentricity attained by the signal during the eye movement search process.

The aggregate $d'$ across all trials can be calculated by weighting each $d'_E$ by an estimate of the probability of the minimum retinal eccentricity of the signal attain a value E, $m_E = p(E_{min})$. The effective $d'$ across trials is then estimated as:

$$\langle d' \rangle = \sum_{E=0}^{e} p(E_{min}) d'_E \qquad (10)$$

In practice, $p(E_{min})$ can be estimated empirically by using an eye tracker during the observers' search. In this paper, we used the eye position measurements to calculate the closest fixation point to the signal location (in signal-present trials) and created a probability histogram of the distance distributions. Figure 1a shows three samples of eye movements from three different observers for the same trial. Solid lines represent saccades and fixations. The white circle is placed at the location of the signal. Using these fixations and the signal location, we can calculate the closest fixation point for each trial (dotted line) and estimate a probability distribution.

### 2.2.2.4 *Estimates of minimum signal eccentricity from time and display size (Time closest fix)*

Explicit knowledge of observers' eye movement behavior requires an eye tracker that is not always available to researchers and technology developers. Thus, we explore a simpler approximation to the distribution of the minimum retinal eccentricities of the signal. This method used the median fixation time from the eye tracker (ET) and the trial search time. We combined these numbers with the stimulus size (in degrees visual angle) to estimate a number of fixations per trial and infer the closest fixation to the signal.

The method first estimates the number of fixations per trial from response time/fixation time. We used the average fixation time, 250 ms in 2D search and 500 ms in 3D search, and the average response times, which were 3.16 seconds for 2D and 22.62 seconds for 3D search. We also considered the number of scrolls per trial for the 3D search, which was 100 slices, divided the number of fixations per number of slices, and assumed a minimum of 1 fixation per slice. We then assumed that the fixations are distributed along equidistant points on a rectangular grid. We subsequently calculated the minimum distance to any fixations for a signal placed at each $x, y$ locations in the image.

$$D(x,y) = \min_{J} \left[ \sqrt{(x - x_{fix,j})^2 + (y - y_{fix,j})^2} \right] \quad (11)$$

where $x_{fix,j}$ and $y_{fix,j}$ refer to the locations of the $j^{th}$ fixation, and min is the minimum function across the $J$ fixations. The probability of each minimum eccentricity can be calculated:

$$m_E = p(\langle E_{min} \rangle) = \frac{1}{wh} \sum_x^w \sum_y^h [D(x,y) = E_{min}] \quad (12)$$

where $h$ and $w$ stand for the height and width of the image and […] are Iverson brackets, defined as [Q] = 1 if Q is true and [Q] = 0 if Q is false.

In practice, we bin minimum retinal eccentricities into various discrete categories resulting in a discrete number of eccentricities. Figure 1b shows a graphic representation of the minimum distance (grey value) as a function of signal position for various equidistant fixations. Dark values represent small distances to the fixation point, while brighter values represent larger distances.

### 2.3 Template Calculation as a function of eccentricity for Foveated Models

A fundamental component in calculating $d'$ as a function of retinal eccentricity is the estimation of the appropriate templates, $\mathbf{w}_E$. Lago et al. [80], used a model in which perceptual templates decrease in spatial resolution with increasing distance away from the point of fixation (retinal eccentricity). For the classic Channelized Hotelling Observer (CHO), this is implemented by changing the channels as a function of eccentricity (foveated CHO, or FCHO). For the Non-Prewhitening model with Eye Filter (NPWE) the process entails varying the eye filter with eccentricity (foveated NPWE, or FNPWE). The 3D component was also constructed by stacking the 2D templates corresponding to each slice of the signal.

### 2.3.1 Foveated Channelized Model Observer (FCHO)

For the FCHO model, we modified the Gabor channels as a function of the distance from the target to the point of fixation. Therefore, at the fovea (0 degrees of eccentricity), we use the original Gabor channels from the CHO standard model: 8 orientations and 6 spatial frequencies (16, 8, 4, 2, 1 and 0.5 cycles per degree). Then, as the distance increases, the size for all the Gabor channels is non-linearly scaled in respect to the eccentricity E in degrees of visual angle.

$$scaling = 1 + \alpha E^{\beta} \qquad (13)$$

where $\alpha=0.7063$, $\beta=1.6953$ and $K=2.7813$. The scaling parameters were optimized to predict performance of the two studied signals as a function of retinal eccentricity (Lago et al., 2019 [80]). The central spatial frequencies of the Gabors decrease inversely to the scaling. That way, spatial frequencies (in cycles/degree) for eccentricity 1 dva would be: 9.3770, 4.6885, 2.3443, 1.1721, 0.5861, and 0.2930. For eccentricity 2 dva: 4.8672, 2.4336, 1.2168, 0.6084, 0.3042, and 0.1521. For eccentricity 3 dva: 2.8837, 1.4419, 0.7209, 0.3605, 0.1802, and 0.0901. And so on for further eccentricities. Additionally, Gabor channels with a frequency smaller than 0.15 cycles/degree are removed from the template due to their size being bigger than the image. Figure 2a shows how channels of a given foveal spatial frequency and orientation scales up with retinal eccentricity. The figure shows all eight orientations but only one spatial frequency.

All 8 orientations are used for all eccentricities. These templates have access to fewer high spatial frequency-tuned channels at higher eccentricities, thus reducing their signal detection accuracy.

### 2.3.2 Foveated Non-Prewhitening Model Observer with Eye Filter (FNPWE)

The foveated extension of the NPWE was implemented by changing the contrast sensitivity function, or eye filter, as a function of the distance from the fixation point to the signal. As the distance increases, the new contrast sensitivity function is calculated with respect to the eccentricity E in degrees of visual angle.

$$\mathcal{E}(\rho) = (\rho E^n)^\alpha \exp(-\beta(\rho E^n)^\gamma) \qquad (14)$$

Values for $\alpha, \beta, \gamma$ and $n$ were optimized to predict the human performance of the two signals as a function of retinal eccentricity ($\alpha$=0.83, $\beta$=0.35, $\gamma$=0.4, and $n$=2.2). Figure 2b shows how the eye filter's sensitivity to spatial frequencies varies with retinal eccentricity. The figure shows three retinal eccentricities and how the model's access to high spatial frequencies diminishes with retinal eccentricity.

### 2.3.3 Adjusting Internal Noise

The final decision variable accounted for additive internal noise sampled from a Gaussian distribution $\epsilon_{int} \sim \mathcal{N}(0, (K\sigma_\lambda)^2)$ which standard deviation was proportional to the standard deviation of the model $\sigma_\lambda$ and adds one fitting parameter K to the model. This parameter was fit to 2.78 for the FCHO model and to 15.13 for the FNPWE model.

### 2.4 Sample-driven Foveated Model Observer with Eye Movements and Scrolls

To validate the image quality metrics for the foveated model observers, we compared its performance prediction to that of the full implementation of the sample-driven Foveated Search Model (FSM). The complete FSM uses the different templates across eccentricities to calculate decision variables, explores the images through eye movements and scrolls, integrates decision variables for each location across the search process, and reaches final decisions about signal presence/absence. Here, we briefly discuss the FSM and refer to Lago et al. 2019 [80] for full details of the model.

The FSM processes the entire image in parallel with the different templates given by the distance of the image region from the point of fixation. The template response at the coordinate, $p$, is calculated by using a template for retinal eccentricity as determined by the distance of the image subregion, $\mathbf{g}_p$, from the current fixation:

$$\lambda_p = \mathbf{w}_E^t \mathbf{g}_p + \epsilon_{int} \tag{15}$$

To construct these templates, we use the modified channelized hotelling observer described in section II.C.1.

Once the templates are built, we assign an explicit fixation point, and we take the template's responses at the corresponding distance (in degrees of visual angle) to the given fixation point. Template responses are transformed to a likelihood ratio. Integration between fixations is made by multiplying the likelihood ratios of each fixation. The final decision takes the highest likelihood ratio for all fixations (and 3D scrolls).

For this paper, the FSM uses human eye movements and 3D scrolls collected during the human observer experiments. The FSM is guided by the list of fixations/scrolls in the corresponding trial for each participant. This way, the model is processing the same fixations that the human observer was fixating.

## 3 Materials and Methods

The $d'$ metrics above were evaluated for a 2D search and 3D search of two signal types with different sizes. The foveated model observers (FCHO and FNPWE) parameters were fit to a separate experiment that measured human detection of the signals as a function of retinal eccentricity in a location known exactly task. Here, we briefly describe the tasks but refer to Lago et al., 2021 [63] for more detail.

### 3.1. Synthetic Noise Stimuli

Stimuli for the synthetic noise field were generated from 3D white noise fields ($\mu = 128, \sigma = 25$) filtered by a power spectrum ($f^{-2.8}$) using frequency indexes. The noise field size was 1024×820×100 voxels.

Two signals were generated: a small spherical sharp sphere (0.13 degrees of visual angle) that we refer as MCALC and a 3D Gaussian blob, that we refer to as MASS (3 standard deviations = 0.66 degrees of visual angle). Stimuli had a 50% chance of having one of the two targets (divided 50:50 between MCALC andMASS). The MASS signal was highly present on the lower range of spatial frequencies along with the background noise. The amplitude of the targets was set to be 83 gray levels over the background mean (128) at their peak.

To generate the 2D trials, we selected the central slice of the signal (most information present), or a random slice on signal-absent trials.

### *3.2. Psychophysical Search Experiment*

Seven undergraduate students at the University of California, Santa Barbara, participated in exchange for course credit. Observers sat at 75cm distance from a 1280x1024 resolution monitor in a darkened room. The monitor luminance was linearly calibrated between ~0 cd/m$^2$ and 111 cd/m$^2$ for gray levels 0 and 255, respectively. At the beginning of each trial, observers were asked to search for a specific target without a time limit. Trials were 2D and 3D intertwined. For the 3D search, observers were able to scroll freely using the mouse. A non-overlapping scroll bar was present on the right side of the screen. When observers pressed the spacebar, a question asking about the presence of the target was shown. Feedback was always given, showing the correct response. A real-time eye tracker (Eyelink 1000, SR Research Inc.) was recording fixations. We used the vendor's default parameters: eye velocity and acceleration thresholds of 30 degrees/sec and 9,500 degrees/sec$^2$, respectively. Figure 3 shows the timeline for one trial of this experiment. Participants were provided informed consent and treated according to the approved human subject research protocol by the University of California, Santa Barbara: 12-18-0025, 12-16-0806, and 12-15-0796.

### *3.3. Psychophysical Location Known Exactly Experiment: d' vs eccentricity*

In order to know how the detectability of both signals decreases at increasing eccentricities, we ran a separate gaze-contingent experiment. Seven human observers participated in this study. Human observers were asked about the presence of a given signal (50% between MCALC or MASS) at a cued location while fixating at different distances. The signal was present in 50% of the trials. An eye tracker monitored that participants were not making an eye movement, in which case, the trial was discarded. Distances measured were 1, 3, and 6 degrees of visual angle for the MCALC and 0, 3, 6, and 9 degrees for the MASS. FCHO and FNPWE parameters were fit to human data from this experiment.

### 3.5. Quantifying goodness of fit of models to human data.

To quantify the ability of the various models to predict human performance we computed a log-likelihood measure (Table 1) using the following formula:

$$LL = -\log[\frac{1}{\sqrt{2\pi\sigma^2}}\exp\left(-\frac{(human - model)^2}{2*stderr^2}\right)] \quad (16)$$

## 2 Results

We first present the result of fitting the FCHO and FNPWE models to the human data for the detection of the two signals (MCALC and MASS) as a function of retinal eccentricity (Figure 4a). A single set of model parameters were used to fit the curves for both signals simultaneously. Results show how detectability falls off more pronouncedly with eccentricity for the MCALC signal compared to the MASS signal.

To integrate the $d'$ indices into a single figure of merit, we considered various methods. Figure 4b shows the various weighting functions considered. The simplest weighting functions are the average $d'_E$ and the normalized $d'_E$ weighted. The other two schemes propose a weighting based on the probability distribution of the closest fixation to the signal (separately shown for the MCALC and MASS signals). The ET-closest fix approach uses actual fixation measurements from the observers to estimate the probabilities. An interesting finding is that the distribution of the closest fixation varies across signal types. For the MASS signal, the distribution of the closest fixation is similar across 2D and 3D search. In contrast, for the MCALC signal, the distribution is narrower for the 2D search and broader for the 3D search, suggesting that observers have difficulty guiding their eye movements and fixating the small signal in the 3D search.

The last method to integrate the various $d'_E$, Time closest fixation, also relies on the closest fixation to the signal but approximates these distributions based on observer decision times and a simplifying assumption about the fixations. Figure 4b (right graph) shows the weighting function for this latter method.

Performance predictions for the 2D and 3D search require taking the dot product of the $d'_E$ function (Figure 4a) and the weighting function (Figure 4b, equation 8). Figure 5 shows human and estimated $d'$ performance using different weighting functions for 2D (left), and 3D (right) search. The top panel corresponds to the metrics based on the FCHO model, while the bottom panel to the FNPWE model. Both panels include human performance as measured by the empirical $d'$ calculated from seven observers. The panels also show the predicted $d'$ using the full implementation of a Foveated Search Model (FSM) that executes the same eye movements and scrolls of each human observer for the corresponding trial [62].

The pattern of results is similar across both model observers (FCHO and FNPWE). The average $d'_E$ metric mispredicts the relative search performance for the signal in 2D search. The $d'_E$ weighted method mispredicts the relative performance for the two signals for 3D search. All metrics based on the closest fixation provided a better approximation of humans' relative performance and the FSM model across conditions and signals. Table 1 shows the negative log-likelihood of observing the human data given each model. A smaller value suggests a better fit. The results confirm that the

## 3 Discussion

Search in three-dimensional images requires observers to use the visual periphery to sample the slices through eye movements and can lead to performance dissociations with 2D search. In particular, small signals, difficult to detect in the visual periphery, result in these dissociations. Our results showing how the distribution of the closest fixation to the signal varies for the MCALC signal in 2D and 3D search illustrates this point. The finding is related to the smaller size and sharper edges of the MCALC, which are progressively filtered out with increasing eccentricity. The high spatial frequency information is particularly important for tasks in $1/f^{2.8}$ noise where the noise decreases with spatial frequency. For 2D search, a great portion of the search space can be covered by the fixational sampling, and results show that observers fixate the signal in 40% of the trials. For the 3D search, consisting of 100 slices, the observers only cover a small fraction of the regions and fixate the signal in only 15% of the trials.

The Foveated Search Model (FSM) shows a strong agreement with human detectability for both MCALC and MASS in both tasks. However, the computational complexity of the full calculation of the Foveated Search Model (FSM) can sometimes reduce its usability. The proposed metrics presented in this paper are based on how the two signals are detected in the periphery (Figure 4a) and how that interacts with their detectability in 2D vs 3D search. The metrics are based on classic expressions for index of detectability commonly used in the medical image quality community [1], [2], [5], [7], [64], [72], [81]

When combining the $d'$ indexes with a simple average, results are not different in 2D and 3D, and estimation is higher for the MASS in both (similar to human results in 3D). Additionally, when using a normalized $d'$ as the weighting function, results are relatively opposites: MCALC shows a higher $d'$ than MASS for both 2D and 3D (similar to human results in 2D).

Finally, using an estimation of closest distances from both eye movements (ET closest) and trial times (Time closest) provides a $d'$ estimation that predicted human results (and FSM results) for both signals in 2D and 3D search. Although this estimation is still far from human results, it can explain possible dissociations in signal detectability in 2D and 3D search that is not captured by standard model observers.

Table 2 summarizes the advantages and disadvantages of each of the methods presented in this paper and standard model observers in the literature [4], [5], [39], [81]–[83].

## 4 Conclusion

As with all modalities, evaluating 3D medical imaging systems is an important step for characterizing their performance and optimizing the many components that define them. However, model observers traditionally used for this purpose might not capture the complexities of human visual search in volumetric images. Foveated models like the FSM can be a solution in this case, but their computational cost is sometimes prohibitive. This paper presented two different ways to overcome this complexity by estimating the closest fixations during trials and combining it with the signal detectability at different retinal eccentricities. The methodologies might not provide quantitative predictions of human performance as accurate as a foveated search model. Still, they can highlight potential dissociations in signal detectability that point to the need for more investigation for a specific imaging system and thus a useful tool for engineers and medical physicists.

*Disclosures*

This research was performed under an IRB protocol for human data (12-16-0806) approved by the University of California, Santa Barbara. The authors declare no conflict of interest.

*Acknowledgments*

The research was funded by the National Institute of Health grants R01 EB018958 and R01 EB026427. We thank the NIH RSNA Perception lab at the RSNA annual meeting for providing an excellent opportunity to conduct psychophysics experiments with radiologist participants.

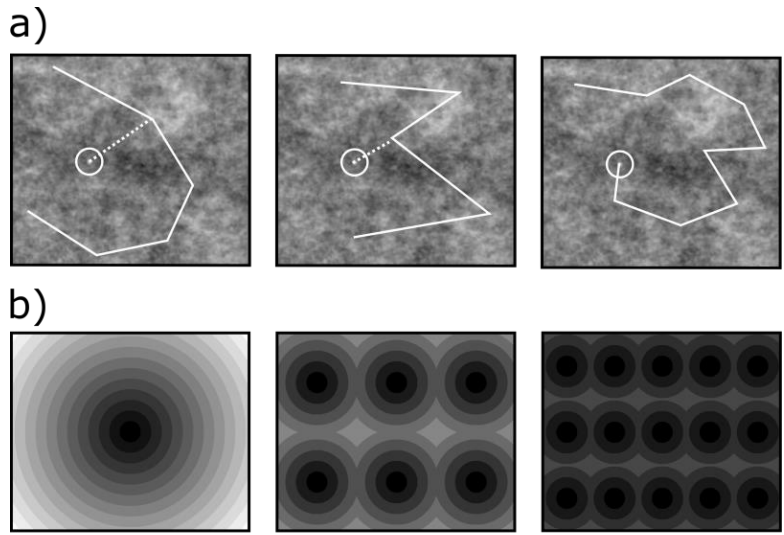

Figure 1. a) Samples of eye movements (solid lines) and target location (circle) used to calculate the closest fixation point (dotted lines) from the eye tracker samples. b) Method to estimate the distribution of minimum signal retinal eccentricities from response time, median fixation time, and display size. Examples are for estimation of 1 fixation, 6 fixations, and 15 fixations. Grey levels stand for the different distances to the closest fixation point for each pixel in the image.

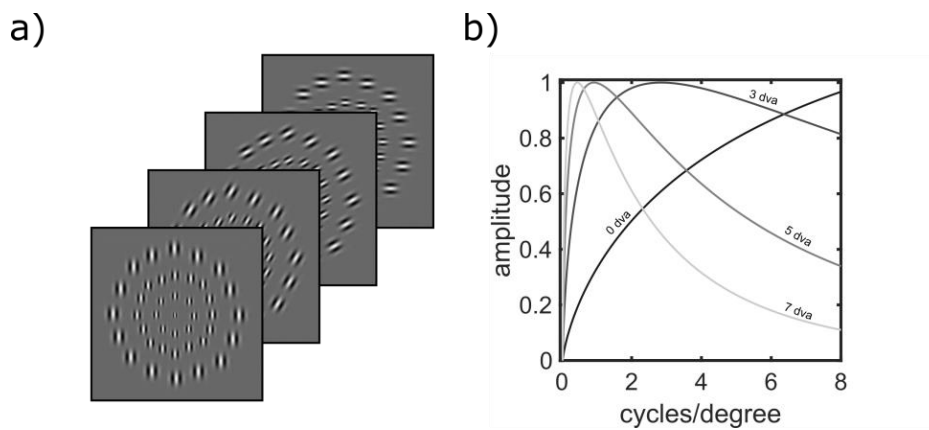

Figure 2: a) Scaling of the Gabor channels for different eccentricities for the foveated channelized Hotelling observer. B) Sample of Contrast Sensitivity Filter (CSF) for different eccentricities for the foveated non-prewhitening model observer with eye filter (right).

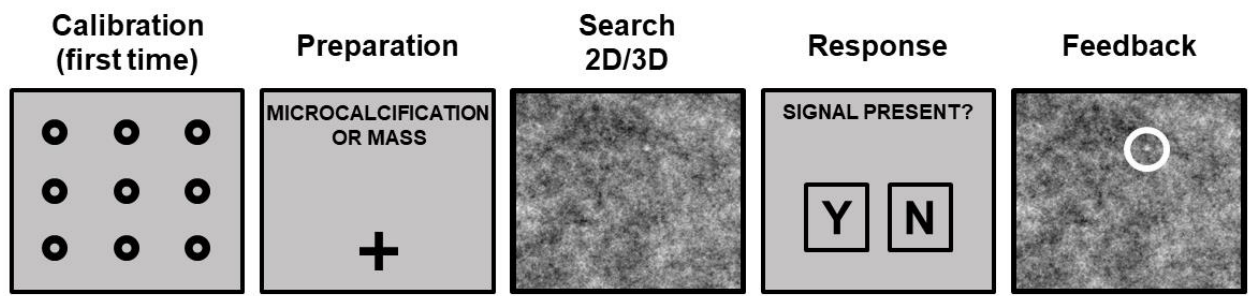

Fig. 3: Timeline of the task for 2D and 3D search for human observers.

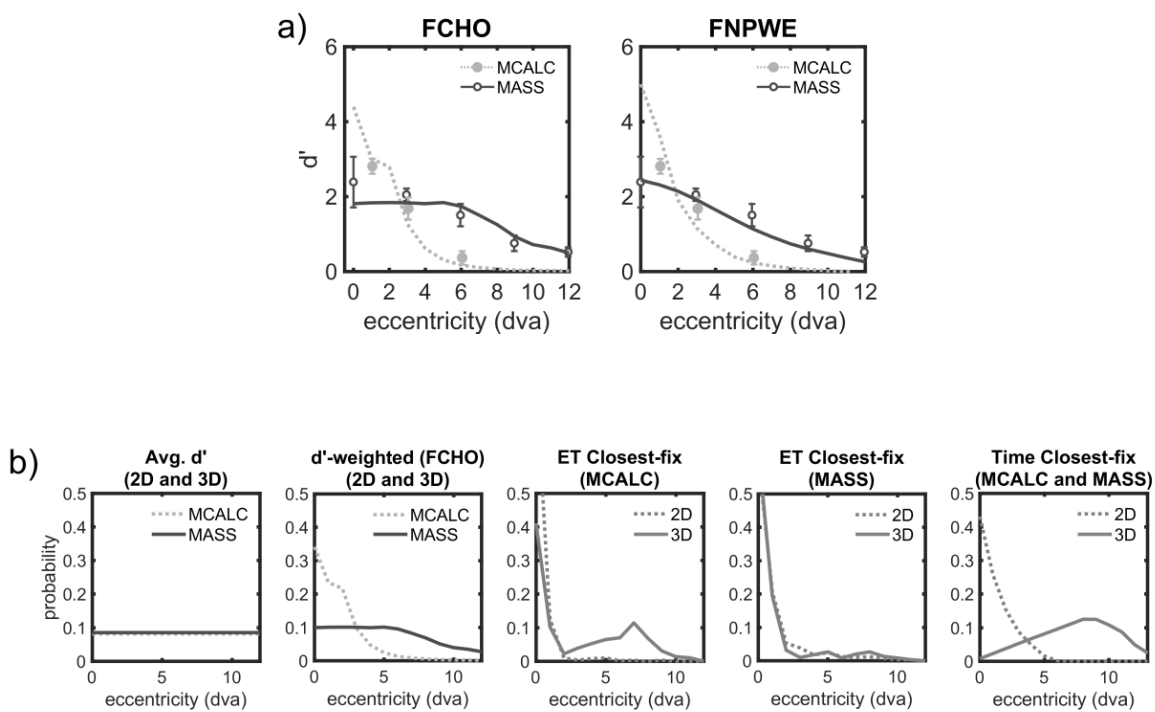

Figure 4. a) Estimated d' from the FCHO and FNPWE models (lines) at each eccentricity (degrees visual angle) along with human perceptual performance (circles) for both signals. b) Each graph is for a different method to aggregate d' across eccentricities into a single figure of merit.

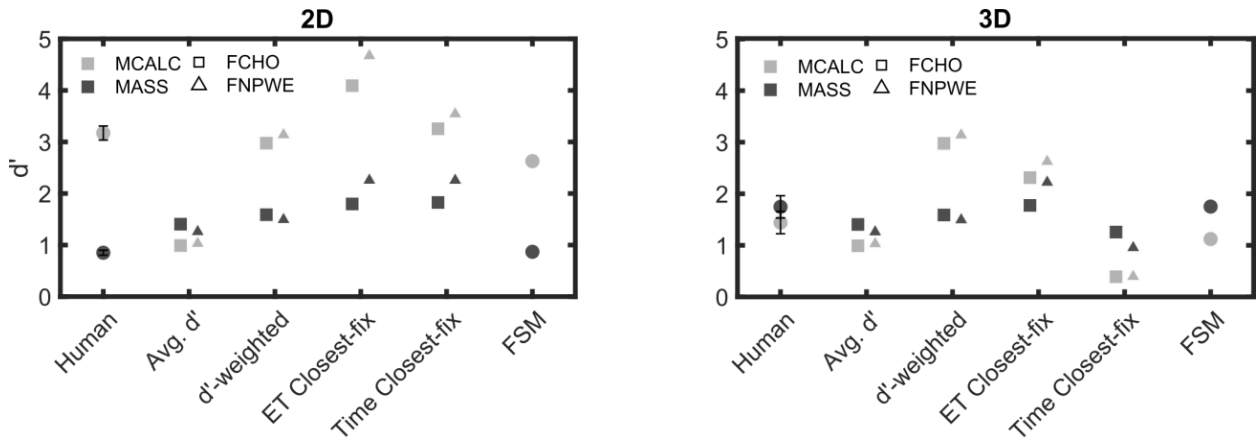

Figure 5: Detectability index for human experiment in 2D (left) and 3D (right) and shorthand calculations (linear, average d', d'-weighted, and closest fixation point) for both microcalcification (MCALC) and mass (MASS). Results are shown for both foveated channelized Hotelling observer (FCHO, squares) and the foveated non-prewhitening model observer with eye filter (FNPWE, triangles).

Table 1: negative log-likelihoods between model observers and human observer data (ratio between MCALC and MASS for each model).

|    | Avg. d' | d'-weighted | ET closest | Time closest | FSM |
|----|---------|-------------|------------|--------------|--------|
| **2D** | 112.79  | 42.88       | 26.40      | 59.35        | 6.15   |
| **3D** | 0.079   | 6.83        | 0.50       | 3.33         | 0.4174 |

Table 2: Advantages and disadvantages for each type of model observers in the literature.

| Conditions | Advantages | Disadvantages |
| --- | --- | --- |
| **Standard Model Observers** | Well-established and computationally simple methodology to assess task performance | Does not consider peripheral processing and erroneously predicts human performance of small signals in 3D search |
| **Foveated Search Model** | Excellent prediction of human performance in 3D search for different signals<br><br>Separately quantifies search and recognition errors | Computationally expensive and requires some knowledge of eye movement explorations using eye trackers |
| **Foveated Model Observer with closest fixation approximation** | Good approximation to the interaction of 3D search, eye movements, and signal's visibility in the visual periphery<br><br>Computationally simpler than foveated search model | Requires an eye tracker to quantify eye movements and determine the closest fixation to signal<br><br>Does not partition errors into search and recognition errors |
| **Foveated Model Observer with decision time approximation** | Computationally simple and does not require eye tracker measurements | Fair approximation to predicting human performance in 3D search for different signals<br><br>Does not partition errors into search and recognition errors |